\documentstyle[12pt,aaspp4]{article}

\def\be{\begin{equation}}
\def\ee{\end{equation}}
\def\ba{\begin{eqnarray}}
\def\ea{\end{eqnarray}}

\def\la{\mathrel{\mathpalette\fun <}}
\def\ga{\mathrel{\mathpalette\fun >}}
\def\fun#1#2{\lower3.6pt\vbox{\baselineskip0pt\lineskip.9pt
        \ialign{$\mathsurround=0pt#1\hfill##\hfil$\crcr#2\crcr\sim\crcr}}}

\begin{document}
\null\vspace{-62pt}
\begin{flushright}
astro-ph/9806185\\
%June 12, 1998
%August 5, 1998
%December 1, 1998
%\today
September 10, 1999\\
to appear in ApJ, 531, \#2 (March 10, 2000)
\end{flushright}
\title{Supernova pencil beam survey}

\author{Yun Wang\footnote{Present address: Dept. of Physics, 
225 Nieuwland Science Hall, University of Notre Dame, Notre Dame, 
IN 46556-5670. email: Yun.Wang.92@nd.edu}}
\affil{{\it Princeton University Observatory} \\
{\it Peyton Hall, Princeton, NJ 08544\\}
{\it email: ywang@astro.princeton.edu}}

\begin{abstract}

Type Ia supernovae (SNe Ia) can be calibrated to be good standard candles
at cosmological distances. We propose a supernova pencil beam survey
that could yield between dozens to hundreds of SNe Ia in redshift bins 
of 0.1 up to $z=1.5$, which would compliment space based SN searches,
and enable the proper consideration 
of the systematic uncertainties of SNe Ia as standard candles, 
in particular, luminosity evolution and gravitational lensing.
We simulate SNe Ia luminosities by adding weak lensing noise (using empirical
fitting formulae) and scatter in SN Ia absolute magnitudes
to standard candles placed at random redshifts.
We show that flux-averaging is powerful in reducing the combined
noise due to gravitational lensing and scatter in SN Ia
absolute magnitudes.
The SN number count is not sensitive to matter distribution 
in the universe; it can be used to test models of cosmology
or to measure the SN rate.
The SN pencil beam survey can yield a wealth of data which
should enable accurate determination of the cosmological parameters
and the SN rate, and provide valuable information on the 
formation and evolution of galaxies.

The SN pencil beam survey can be accomplished on a dedicated 4 meter 
telescope with a square degree field of view. 
This telescope can be used to conduct 
other important observational projects compatible with the SN pencil beam 
survey, such as QSOs, Kuiper belt objects, and in particular,
weak lensing measurements of field galaxies, and the search for gamma-ray 
burst afterglows.

\end{abstract}

%\keywords{Cosmology}

\section{Introduction}

Toward the end of the millennium, cosmology has matured into
a phenomenological science. Observational data now dominates
aesthetics in the evaluation of cosmological models.
Of fundamental importance is the determination of cosmological
parameters, in particular, the Hubble constant $H_0$, the
matter density faction $\Omega_m$, and the density fraction contributed
by the cosmological constant $\Omega_{\Lambda}$.
Observation of distant Type Ia supernovae (SNe Ia) has become an
increasingly powerful means of measuring cosmological parameters
(\cite{Perl97}, 1998; \cite{Riess98}; \cite{Schmidt98}),
because SNe Ia can be calibrated to be good standard candles
at cosmological distances. (\cite{Riess95})

A type Ia SN is the thermonuclear explosion of a carbon-oxygen white
dwarf in a binary when the rate of the mass transfer from the companion
star is high. The SN explosion blows the white dwarf completely apart. 
The radioactive decay of the isotopes $^{56}$Ni
and $^{56}$Co is responsible for much of the light emitted.
The SN lightcurve reaches a maximum about 15 days after the explosion and 
then declines slowly over years.
A SN can outshine the galaxy in which it lies. 

Two independent groups (\cite{Riess98,Perl99}) have 
made systematic searches for SNe Ia for the purpose of
measuring cosmological parameters. Their preliminary results 
seem to indicate a low matter density universe, possibly with a 
sizable cosmological constant. Even though the uncertainty in
these results is large, they clearly demonstrate that 
the observation of SNe Ia can potentially become a reliable 
probe of cosmology. 
However, there are important systematic uncertainties of SNe Ia as 
standard candles, in particular, luminosity evolution and 
gravitational lensing.
To constrain the evolution of SN Ia peak absolute luminosities, we need a large
number of SNe Ia at significantly different redshifts (low $z$ and $z>1$),
which is not available at present.
Both groups have assumed a smooth universe
in their data analysis, although they include
lensing in their error budgets. 
Since we live in a clumpy universe,
the effect of gravitational lensing must be taken into account
adequately for the proper interpretation of SN data.
At present, the small number of observed high $z$ SNe Ia 
prevents adequate modeling of gravitational lensing effects.
  
In this paper, we propose a pencil beam survey of SNe Ia 
that could yield between dozens to hundreds of SNe Ia in redshift bins 
of 0.1 up to $z=1.5$ (see \S 3) which would allow the proper modeling of
gravitational lensing, as well as a quantitative understanding of 
luminosity evolution.
Such a survey would yield a wealth of data which can be used
to make accurate measurement of cosmological parameters
and the SN rate, and provide
powerful constraints on various aspects of the cosmological model.
In \S 2, we consider the weak lensing of SNe Ia.
We simulate SN Ia luminosities by adding weak lensing noise
and scatter in SN Ia absolute magnitudes
to standard candles placed at random redshifts.
We show how flux-averaging reduces the combined noise of gravitational
lensing and scatter in SN Ia absolute magnitudes.
In \S 3, we show that the number count of SNe 
from a pencil beam survey is not sensitive to matter distribution in 
the universe; it can be used as a 
test of models of cosmology and SN progenitors, 
or to measure the SN rate accurately.
In \S 4, we discuss the observational feasibility of the SN pencil beam survey.
\S 5 contains conclusions.

\section{Weak lensing of supernovae}

In a SN Ia Hubble diagram, one must use distance-redshift relations
to make theoretical predictions. Unlike angular separations
and flux densities, distances are not directly measurable, but they
are indispensable theoretical intermediaries.
The distance-redshift relations depend on the distribution of
matter in the universe. 

In a smooth Friedmann-Robertson-Walker (FRW) universe,
the metric is given by $ds^2=dt^2-a^2(t)[dr^2/(1-kr^2)+r^2 (d\theta^2
+\sin^2\theta \,d\phi^2)]$, where $a(t)$ is the cosmic scale factor,
and $k$ is the global curvature parameter ($\Omega_k
=1-\Omega_m-\Omega_{\Lambda}=-k/H_0^2$).
The comoving distance $r$ is given by
\be
\label{eq:r(z)}
r(z)=\frac{cH_0^{-1}}{|\Omega_k|^{1/2}}\,
\mbox{sinn}\left\{ |\Omega_k|^{1/2}
\int_0^z dz'\,\left[ \Omega_m(1+z')^3+\Omega_{\Lambda}+\Omega_k(1+z')^2
\right]^{-1/2} \right\},
\ee
where ``sinn'' is defined as sinh if $\Omega_k>0$, and sin if  $\Omega_k<0$.
If $\Omega_k=0$, the sinn and $\Omega_k$'s disappear from Eq.(\ref{eq:r(z)}),
leaving only the integral. 
The angular diameter distance is given by $d_A(z)=r(z)/(1+z)$,
and the luminosity distance is given by $d_L(z)=(1+z)^2 d_A(z)$.

However, our universe is clumpy rather than smooth.
According to the focusing theorem in gravitational lens theory,
if there is any shear or matter along a beam connecting 
a source to an observer, the angular diameter distance of
the source from the observer is {\it smaller} than that which would occur
if the source were seen through an empty, shear-free cone,
provided the affine parameter distance (defined such that its element
equals the proper distance element at the observer) 
is the same and the beam has not gone through a caustic.
An increase of shear or matter density along the beam decreases the
angular diameter distance and consequently increases the
observable flux for given $z$. (Schneider, Ehlers, \& Falco 1992)

The observation of SNe Ia at $z >1$ is important for the determination
of cosmological parameters, but the dispersion in SN Ia luminosities 
due to gravitational lensing can become comparable to
the intrinsic dispersion of SNe Ia absolute magnitudes because
the optical depth for gravitational lensing increases with redshift.

\subsection{Direction dependent smoothness parameter}

If only a fraction $\tilde{\alpha}$ (known as
the smoothness parameter) of the matter density is smoothly
distributed, the largest possible distance (for given redshift) for
light bundles which have not passed through a caustic is given by
the appropriate solution to the following equation:
\be
\label{eq:DR}
g(z) \, \frac{d\,}{dz}\left[g(z) \frac{dD_A}{dz}\right]
+\frac{3}{2} \tilde{\alpha} \,\Omega_m (1+z)^5 D_A=0,
\ee
where $g(z) \equiv (1+z)^3 \sqrt{ 1+ \Omega_m z+ \Omega_{\Lambda}
[(1+z)^{-2} -1] }$. (\cite{Kantow98})
The $\Omega_{\Lambda}=0$ form of Eq.(\ref{eq:DR}) has been known as
the Dyer-Roeder equation. (\cite{DR73,Sch92})

Fig.1(a) shows magnitude versus redshift for the three cosmological models
considered by Riess et al. (1998),
SCDM ($\Omega_m=1$, $\Omega_{\Lambda}=0$), 
OCDM ($\Omega_m=0.2$, $\Omega_{\Lambda}=0$),
and $\Lambda$CDM ($\Omega_m=0.2$, $\Omega_{\Lambda}=0.8$).
For each cosmological model, the upper curve represents the completely
clumpy universe (empty beam, $\tilde{\alpha}=0$), while the lower curve 
represents the completely smooth universe (filled beam,
$\tilde{\alpha}=1$). 
Fig.1(b) shows the same models relative to smooth OCDM (
filled beam, $\tilde{\alpha}=1$),
the middle curve for each model now represents a universe with half
of the matter smoothly distributed (half-filled beam, $\tilde{\alpha}=0.5$).
Clearly, at $z>1$, there is degeneracy of distances in
a flat clumpy universe and an open smooth universe,
and also in an open clumpy universe and a flat smooth universe
with a sizable cosmological constant,
as has been noted by a number of previous authors.
(\cite{Kantow98,Linder98,Holz98b})

We can generalize the angular diameter distance $D_A(z)$
by allowing the smoothness parameter $\tilde{\alpha}$
to be {\it  direction  dependent}, i.e., a property of the {\it beam} 
connecting the observer and the standard candle.
The smoothness parameter $\tilde{\alpha}$ essentially represents
the amount of matter that causes weak lensing of a given source.
Since matter distribution in our universe is inhomogeneous, we can think
of our universe as a mosaic of cones centered on the observer, 
each with a different value of $\tilde{\alpha}$. 
This reinterpretation of $\tilde{\alpha}$ implies
that we have $\tilde{\alpha}>1$ in regions of the universe in which
there are above average amounts of matter which can cause magnification
of a source. (\cite{Wang99a})

In order to derive a unique mapping between the distribution
in distances and the distribution in the direction dependent 
smoothness parameter for given redshift $z$, we {\it define} 
the direction dependent smoothness parameter
$\tilde{\alpha}$ to be the solution of Eq.(\ref{eq:DR})  
for given distance $D_A(z)$.

At given redshift $z$, the magnification of a source can be expressed 
in terms of the apparent brightness of the source 
$f(\tilde{\alpha}|z)$, or in terms of the angular diameter distance 
to the source $D_A(\tilde{\alpha}|z)$:
\be
\label{eq:mu}
\mu = \frac{ f(\tilde{\alpha}|z)}{ f (\tilde{\alpha}=1|z)}
= \left[ \frac{D_A(\tilde{\alpha}=1|z)}{D_A(\tilde{\alpha}|z)} \right]^2,
\ee
where $f (\tilde{\alpha}=1|z)$ and $D_A(\tilde{\alpha}=1|z)$
are the flux of the source and angular diameter distance 
to the source in a completely smooth universe (filled beam), 
and $\tilde{\alpha}$ is the direction dependent smoothness parameter.
Since distances are not directly measurable, we should interpret 
Eq.(\ref{eq:mu}) as defining a unique mapping between the magnification
of a standard candle at redshift $z$ and the direction dependent 
smoothness parameter $\tilde{\alpha}$ at $z$; $\tilde{\alpha}$ parametrizes 
the direction dependent matter distribution in a well-defined manner.

From the magnification distributions of standard candles 
at various redshifts, $p(\mu|z)$, 
with $z=$0.5, 1, 1.5, 2, 2.5, 3, 5, found numerically by
Wambsganss et al. for $\Omega_m=0.4$, $\Omega_{\Lambda}=0.6$ 
(\cite{Wamb97,Wamb99}),
Wang (1999a) has obtained simple empirical fitting formulae 
for the distribution of $\tilde{\alpha}$:
\be
\label{eq:p(alpha)}
p(\tilde{\alpha}|z)=C_{norm}\, \exp\left[ -\left( \frac{\tilde{\alpha}-
\tilde{\alpha}_{peak}}
{w \,\tilde{\alpha}^q} \right)^2 \right],
\ee
where $C_{norm}$, $\tilde{\alpha}_{peak}$, $w$, and $q$ depend on $z$ 
and are independent of $\tilde{\alpha}$. They are given by
\ba
\label{eq:aq}
C_{norm}(z) &=&   10^{-2} \left[   0.53239
  + 2.79165 \, \left(\frac{z}{5}\right)
  - 2.42315\, \left(\frac{z}{5}\right)^2
  + 1.13844\,\left(\frac{z}{5}\right)^3 \right], \nonumber \\
\tilde{\alpha}_{peak}(z) &=&     1.01350    
   -1.07857  \, \left(\frac{1}{5 z}  \right)
   +2.05019  \, \left(\frac{1}{5z}\right)^2  
   -2.14520  \, \left(\frac{1}{5z}\right)^3, \nonumber  \\
w(z) &=&    0.06375
   + 1.75355  \, \left(\frac{1}{5 z} \right) 
   - 4.99383    \, \left(\frac{1}{5z}\right)^2
   + 5.95852    \, \left(\frac{1}{5z}\right)^3, \nonumber \\
q(z) &=&    0.75045    
   +1.85924 \, \left(\frac{z}{5}   \right)
   -2.91830 \, \left(\frac{z}{5}\right)^2   
   +1.59266 \,\left(\frac{z}{5}\right)^3.
\ea   
$C_{norm}(z)$ is the normalization constant for given $z$.
The parameter $\tilde{\alpha}_{peak}(z)$ indicates the average 
smoothness of the universe at redshift $z$, it increases with $z$
and approaches $\tilde{\alpha}_{peak}(z)=1$ (filled beam) at $z=5$;
the parameter $w(z)$ indicates the width of the distribution in
the direction dependent smoothness parameter $\tilde{\alpha}$,
it decreases with $z$.
The $z$ dependences of $\tilde{\alpha}_{peak}(z)$ and $w(z)$ 
are as expected because as we look back to earlier times, 
lines of sight become more filled in with matter, and
the universe becomes smoother on the average.
The parameter $q(z)$ indicates the deviation of $p(\tilde{\alpha}|z)$
from Gaussianity (which corresponds to $q=0$).

Models with different cosmological parameters should lead to
somewhat different matter distributions $p(\tilde{\alpha}|z)$.
In the context of weak lensing of 
standard candles, we expect the cosmological parameter dependence 
to enter primarily through the magnification $\mu$ to direction 
dependent smoothness parameter $\tilde{\alpha}$ mapping at given $z$ 
(the same $\tilde{\alpha}$ corresponds to very different $\mu$ in 
different cosmologies).

\subsection{Flux averaging of SN luminosities}

Gravitational lensing noise in the Hubble diagram can be reduced 
by appropriate flux averaging of SNe Ia in each redshift bin.
Because of flux conservation, the average flux of a sufficient number
of SNe Ia at the same $z$ from the same field should be the same as
the true flux of the SNe Ia without gravitational lensing
if the sample is complete.

It is convenient to compare the distance modulus of SNe Ia, $\mu_0$,
with the theoretical prediction
\be
\mu_0^p= 5\,\log\left( \frac{ d_L}{\mbox{Mpc}} \right)+25,
\ee
where $d_L(z)$ is the luminosity distance.
Before flux-averaging, we convert the distance modulus 
$\mu_0(z_i)$ of SNe Ia into 
``fluxes'', $f(z_i)=10^{-\mu_0(z_i)/2.5}$. We then obtain ``absolute 
luminosities'', \{${\cal L}(z_i)$\}, by
removing the redshift dependence of the ``fluxes'', i.e.,
\be
{\cal L}(z_i) = 4\pi\,d_L^2(z_i|H_0,\Omega_m, 
\Omega_{\Lambda})\,f(z_i),
\ee
where $(H_0,\Omega_m, \Omega_{\Lambda})$ 
are the best-fit cosmological parameters derived from the unbinned data
set \{$f(z_i)$\}. 
We then flux-average over the 
``absolute luminosities'' \{${\cal L}_i$\} in each redshift bin.
The set of best-fit cosmological parameters derived from the binned data 
is applied to the unbinned data \{$f(z_i)$\} to obtain a new set 
of ``absolute luminosities'' \{${\cal L}_i$\}, which is then flux-averaged 
in each redshift bin, and the new binned data is used to derive a 
new set of best-fit cosmological parameters. This procedure is repeated 
until convergence is achieved. 
This iteration should lead to the optimal
removal of gravitational lensing noise
and the accurate determinations of the cosmological parameters.
Wang (1999b) has applied this method to analyze the combined data from the
two groups (\cite{Riess98,Perl99}).

To illustrate how flux-averaging can reduce the dispersion in
SN Ia luminosities caused by weak lensing, let us simulate the data by
drawing $N_{SN}$ random points from the redshift interval $[z_1^0,z_2^0]$,
each point represents a standard candle. We add weak lensing noise by 
giving each standard candle a direction dependent smoothness parameter
$\tilde{\alpha}$ (corresponding to $\mu=
\left[ D_A(\tilde{\alpha}=1)/D_A(\tilde{\alpha}) \right]^2$)
drawn at random from the distribution $p(\tilde{\alpha}|z)$
given in \S 2.1. The scatter in SN Ia
absolute magnitudes, $\Delta m_{abm}$, can be written as
\be
\Delta m_{abm}=\Delta m_{int}+\Delta m_{obs},
\ee
where $\Delta m_{int}$ is the intrinsic scatter and $\Delta m_{obs}$
is the observational noise.
We assume that both intrinsic scatter and observational noise are
Gaussian distributed in magnitude, with dispersions $\sigma_{int}$
and $\sigma_{obs}$ respectively. Then the total scatter in SN Ia
absolute magnitudes is also Gaussian distributed, i.e.,
\be
p(\Delta m_{abm})=\frac{1}{\sqrt{2\pi}\,\sigma_{abm}}\,
\exp\left[ - \frac{(\Delta m_{abm})^2}{2 \sigma_{abm}^2 } \right],
\ee
with $\sigma_{abm}=\sqrt{\sigma_{int}^2+\sigma_{obs}^2}$.
We take $\sigma_{abm}=0.20$. The absolute luminosity of each SN Ia 
extracted from the data is
\ba
{\cal L}(z)&=&10^{-\Delta m_{obs}/2.5}\,\mu(\tilde{\alpha}|z)\,{\cal L}_{int}(z),\nonumber\\
&=& 10^{-\Delta m_{obs}/2.5}\,\mu(\tilde{\alpha}|z) 
\left\{ 10^{-\Delta m_{int}/2.5}\,  
{\cal L}(\tilde{\alpha}=1|z)\right\}\nonumber\\
&=& {\cal L}(\tilde{\alpha}=1|z)\,\mu(\tilde{\alpha}|z)\,
10^{-\Delta m_{abm}/2.5} .
\ea
We have used ${\cal L}_{int}(z)=10^{-\Delta m_{int}/2.5}\,
{\cal L}(\tilde{\alpha}=1|z)$.

For each SN Ia, the total noise is
\be
\Delta m = -2.5 \, \log \left(\frac{ {\cal L}(z)} 
{{\cal L}(\tilde{\alpha}=1|z)} \right).
\ee
Let us average the fluxes of all SNe Ia in the redshift bin $[z_1^0,z_2^0]$:
\be
\overline{\cal L} = \frac{1}{N_{SN}}\, \sum_{i=1}^{N_{SN}} {\cal L}(z_i).
\ee
The flux averaged noise is
\be
(\Delta m )_{avg}= -2.5 \, \log \left(\frac{ \overline{\cal L}} 
{{\cal L}(\tilde{\alpha}=1|\overline{z})} \right),
\ee
where 
\be
\overline{z}=\frac{\sum_{i=1}^{N_{SN}} z_i} {N_{SN}}.
\ee

Table 1 lists the means and dispersions (in the form of mean$\pm$dispersion)
of $\Delta m$ (which are
$\langle \Delta m\rangle$ and $\sigma=\sqrt{
\langle \left[\Delta m-\langle \Delta m\rangle\right]^2\rangle}\,$),
and $(\Delta m )_{avg}$ (which are $\langle(\Delta m )_{avg}\rangle$ and 
$\sigma_{avg}=\sqrt{\langle \left[(\Delta m)_{avg}-\langle (\Delta m)_{avg}
\rangle\right]^2\rangle}\,$)
for various redshift bins with $N_{SN}$=2, 4, and 9 
SNe Ia in each bin, for $10^4$ random samples.
We have taken $\Omega_m=0.4$, $\Omega_{\Lambda}=0.6$.
\begin{table}[htb]
\begin{center}
\caption{Means and dispersions of $\Delta m$ and $(\Delta m )_{avg}$
for a $\Lambda$CDM model.}
\begin{tabular}{cccccc}
\tableline\tableline
$z$ & $\langle \Delta m\rangle\pm \sigma$ 
& $N_{SN}$=2 & $N_{SN}$=4 & $N_{SN}$=9\\
\tableline
$[0.5, 0.6]$ & 0.001 $\pm 0.200 $ 
& -0.007$\pm 0.141$& -0.012 $ \pm 0.101 $& -0.015$\pm 0.067$ \\

$[1, 1.1]$ &  $ 0.002 \pm 0.204 $
& -0.007 $\pm $ 0.145 & -0.013 $\pm $ 0.103 & -0.015 $\pm $ 0.069 \\

$[1.5, 1.6]$ &  0.003$ \pm  $0.210
& -0.006$\pm $ 0.149 & -0.012 $\pm $ 0.107 & -0.015 $\pm $ 0.071 \\

$[1.5, 2]$ & 0.003$ \pm  $0.213
& -0.006$\pm $ 0.151 & -0.012 $\pm $ 0.108 & -0.015 $\pm $ 0.072 \\

$[2, 2.5]$  & 0.005 $ \pm 0.218 $
& -0.005 $\pm $0.155 & -0.012 $\pm $ 0.111 & -0.015 $\pm $ 0.074\\

$[2.5, 3]$   & 0.006 $ \pm  $0.223
& -0.005$\pm $0.159 & -0.012$\pm $0.114 & -0.015$\pm $0.076 \\

$[3, 3.5]$   & 0.007 $ \pm  $0.228
& -0.004$\pm $ 0.162 & -0.012$\pm $0.117 & -0.015$\pm $ 0.078 \\

$[3.5, 4]$ & 0.007 $ \pm 0.233 $
& -0.004$\pm $0.166 & -0.012$\pm $0.119 &-0.015 $\pm $ 0.080 \\

$[4,4.5]$   & 0.008 $\pm $0.238
& -0.004$\pm $0.170 & -0.012$\pm $0.122 & -0.015$\pm $ 0.082\\

$[4.5,5]$  & 0.008$ \pm $0.243
& -0.004$\pm $ 0.174 & -0.012$\pm $ 0.125& -0.016$\pm $ 0.084\\
\tableline
\end{tabular}
%\tablecomments{}
\end{center}
\end{table}

The dispersion decreases roughly as $1/\sqrt{N_{SN}}$.
The dispersion would reduce by 30\% if the sample contains
2 SNe Ia; 50\% if the sample contains 4 SNe Ia.
Even though gravitational lensing noise increases
with redshift, the combined gravitational lensing and SN Ia absolute
magnitude scatter noise 
in the redshift interval $z=[4.5,5]$ can be reduced to the
same level as in the absence of lensing by flux averaging over
two SNe Ia. 

Note that the flux averaged luminosities are biased
towards slightly higher luminosities.
This is as expected, because we have assumed that the intrinsic scatter 
and observational noise in the SN Ia absolute magnitude are Gaussian in 
{\it magnitude}. It is straightforward to show
that the mean of $10^{-\Delta m_{abm}/2.5}$ is 
$\exp\left[(\ln 10/2.5)^2\,\sigma_{abm}^2 /2\right]$, 
which corresponds to a bias of $-\sigma_{abm}^2\, 
\ln 10 /5\simeq -0.018$ for $\sigma_{abm}=0.2$.
If we assume that the intrinsic scatter and observational noise are
Gaussian in luminosity, the flux averaged luminosities become
unbiased.

\section{Supernova number count}

The SN number count is most sensitive to the 
SN rate as a function of $z$, which depends on
the specific rate of SNe, as well as 
the number density of galaxies and their luminosity distribution.
The frequency of SNe is a key parameter for describing
the formation and the evolution of galaxies;
the winds driven by SNe tune the energetics, and their
production of metals determines the chemical evolution
of galaxies and of clusters of galaxies (\cite{Ferrini93,Renzini93}).
The SN II rate is related (for a given initial mass function)
to the instantaneous stellar birthrate of massive stars
because SNe II have short-lived progenitors;
the SN Ia rate follow a slower evolutionary track, and can be used to
probe the past history of star formation in galaxies.
Accurate measurements of the SN rates at intermediate
redshifts are important for understanding
galaxy evolution, cosmic star formation rate and 
the nature of SN Ia progenitors.
(\cite{Madau98a,Ruiz98,Sadat98,Yung98,Madau98b})

The SN rates are very uncertain at present due to the small number 
of SNe discovered in systematic searches.
(\cite{Van94,Pain96}). 
Kolatt \& Bartelmann (1997) have estimated the SN Ia average rate 
per proper time unit per comoving volume to be
\be
\label{eq:SNrate}
n_{SN}(z)=\left[A+Bz\right] \,(100\,\mbox{yr})^{-1}
(h^{-1}\mbox{Mpc})^{-3}, \hskip 1cm A=0.0136, B=0.067,
\ee
for $q_0=0.5$. Changing $q_0$ to lower values leads to lower 
comoving densities but higher luminosities of galaxies at high $z$
(\cite{Lilly95}). Eq.(\ref{eq:SNrate}) has been derived
assuming that all SNe reside in galaxies, neglecting
the redshift dependence of the specific SN rate (number per unit 
luminosity per unit time), and
using the number density of galaxies (as function of redshift)
and the Schechter function parameters derived 
from the Canada France redshift survey (\cite{Lilly95}),
the APM survey (\cite{Loveday92}), and the AUTOFIB survey (\cite{Ellis96}).
We use Eq.(\ref{eq:SNrate}) for all
cosmological models considered in this paper, because it is a very crude 
and conservative estimate, and we use it for the purpose of illustration
only.

The expected number of SNe Ia in a field of angular area $\theta^2$ 
for an effective observation duration of $\Delta t$ up to
redshift $z$ is
\ba
N(z)&=&  \theta^2 \int^z_0 \frac{dr_p(z')}{a(t')}\,
r^2(z')\, \frac{n_{SN}(z')}{(1+z')},\\
&=& 22.4 \,\left(\frac{\theta}{1'}\right)^2 \left(\frac{\Delta t}
{1 \,\mbox{yr}}\right) \int_0^{z}dz'\,
\left[ \frac{r(z')}{cH_0^{-1}}\right]^2\,
\frac{0.0136+0.067z'}{(1+z') \sqrt{\Omega_m(1+z')^3+\Omega_{\Lambda}
+\Omega_k (1+z')^2} }, \nonumber
\ea
where $dr_p=-cdt$ is the proper distance interval, $r$ is comoving 
distance, and the factor $(1+z)^{-1}$ accounts for the cosmological
time dilation.
Note that the number counts fundamentally probe a different
aspect of the global geometry of the universe than do the distance measures
(\cite{Carroll92}).

Fig.2 shows the number of SNe Ia expected per 0.1 redshift interval 
as function of redshift for the same three cosmological models
as in Fig.1,
SCDM ($\Omega_m=1$, $\Omega_{\Lambda}=0$), 
OCDM ($\Omega_m=0.2$, $\Omega_{\Lambda}=0$),
and $\Lambda$CDM ($\Omega_m=0.2$, $\Omega_{\Lambda}=0.8$).
For a one square degree field, and an 
effective observation duration of one year,
the total numbers of expected SNe Ia are 464, 899, and 1705
for SCDM, OCDM, and $\Lambda$CDM respectively for $z$ up to 1.5.

The number of SNe which are strongly lensed by galaxies is given by
\ba
N_{lensed}(z)&=& \int^z_0 dz'\, \tau(z') \, \frac{dN}{dz'},\nonumber\\
&=& 3.81\times 10^{-2} \left(\frac{\theta}{1'}\right)^2 
\left(\frac{F}{0.05}\right) \left(\frac{\Delta t}
{1 \,\mbox{yr}}\right)   \nonumber \\
&& \hskip 1cm \cdot \int_0^{z}dz'\,
 \left[ \frac{r(z')}{cH_0^{-1}}\right]^5
\frac{0.0136+0.067z'}{(1+z') \sqrt{\Omega_m(1+z')^3+\Omega_{\Lambda}
+\Omega_k (1+z')^2} }. 
\ea
We have used the optical depth for gravitational lensing by galaxies
(\cite{Turner90,Fuku91})
\be
\tau(z)\simeq \frac{F}{30} 
\left[ \frac{(1+z) d_A(z)}{cH_0^{-1}}\right]^3,
\ee
where $F$ parametrizes the gravitational lensing effectiveness
of galaxies (as singular isothermal spheres). 
Fig.3 shows the number of strongly lensed SNe as function of
survey depth $z$, for the same cosmological models as in Fig.1 and Fig.2.
For a one square degree field, and an effective 
observation duration of one year,
the total numbers of strongly lensed SNe Ia are 0.2, 0.6, and 1.8 
for SCDM, OCDM, and $\Lambda$CDM respectively for $z$ up to 1.5.

Fig.2 and Fig.3 show that the SN number counts from a pencil beam survey can
be used to measure the SN rate at high redshifts and perhaps to
probe cosmology. Because of the large number of SNe
in each redshift bin and the smallness of gravitational lensing
optical depth, the SN number count should be insensitive
to matter distribution in the universe, and should therefore
provide a robust probe of the cosmological model.
Note that the strongly lensed SNe can be easily removed from 
the survey sample, because they appear as unusually bright SNe.
The number of strongly lensed SNe is very sensitive to the cosmological
model and might be used to further constrain the cosmological model.
But given the small number of strongly lensed SNe Ia expected
in any realistic observational program, their usefulness may be limited.

The SN number counts provide a combined measure of the cosmological 
parameters and the SN rate.  
Fig. 4 shows the parameter dependence of the SN Ia number count per 0.1 
redshift interval ($z-0.05,z+0.05$) as function of redshift $z$. 
Note that the dependences of the SN number count on the
SN rate (parametrized by $A$ and $B$) and the cosmological parameters
are not degenerate and can be differentiated in principle.
In practice, we are ignorant of the functional form of the SN rate
as a function of $z$. Hence,
we may apply the measurements 
of the cosmological parameters as priors to the SN number counts
to obtain a direct measure of the SN rate in the universe
for $z$ up to 1.5, which can be used as a powerful constraint on
the cosmological model.
For a one square degree field, and an effective 
observation duration of one year,
the SN rate per 0.1 redshift interval can be determined to
14-16\%, 9-11\%, 7-8\% at $1<z<1.5$ for SCDM, OCDM, and 
$\Lambda$CDM respectively.

%\newpage
\section{Observational feasibility}

A pencil beam survey would be efficient in the discovery of
SNe at $z\ga 1$ through the combination of data from successive
nights and the comparison of the latest frame of images with all 
previous frames.

SNe Ia are quite faint at $z\sim 1.5$, with AB magnitude in the I band
of $I_{AB}\sim 26$. Because of the UV suppression due to line blanketing
and the apparent IR suppression in the rest frame SN Ia spectrum, 
one should use a passband that corresponds to the wavelength range
of 3000$\AA-10000\AA$ in the SN rest frame; this means
using the I, J, H, or K passbands to observe the $z>1$ SNe.
SN searches from space (HST and NGST) are limited by
small fields of view, a large scale SN search is only possible from the
ground, where one is limited to the I band by the atmosphere.
Here we discuss the observational feasibility of a pencil beam survey
of SNe Ia up to $z\sim 1.5$ in terms of exposure times in the I band;
although multiple band photometry should be obtained to constrain
extinction and evolution of the SNe.

Using values for the photometric parameters from the Sloan Digital
Sky Survey (1 arcsec seeing, effective sky brightness of
20.3 mag/arcsec$^2$ in the I band, seeing-dominated PSF, etc), 
we find that the exposure time for a point source with
AB magnitude of $I_{AB}$ can be written as
\be
t = 13.94 \,{\rm hours} \, 
\left(\frac{S/N}{10}\right)^2 \left(\frac{4\,{\rm m}}
{D}\right)^2 \, 10^{0.8 (I_{AB}-26)}
\ee
where $S/N$ is the signal-to-noise ratio, $D$ is the aperture of
the telescope. The above equation shows that the supernova pencil beam
survey can be accomplished from a modest dedicated 4 meter telescope, 
which can image the same fields (two pencil beams should be observed
to keep the fields close to zenith) every night, which
can lead to the discovery of SNe Ia up to $z=1.5$
via appropriate combination of data from successive nights,
and the light curves of SNe Ia at $z < 1.5$.
It should be feasible to monitor
the faintest SNe Ia from the Keck 10 meter telescope or the HST.

\newcounter{ctr}
\setcounter{ctr}{2}
Spectra of SNe are required to determine whether they are type Ia.
SNe Ia have a Si \Roman{ctr} absorption line at $\sim 4000\,\AA$ that
may be used to identify the $1<z\la 1.5$ SNe Ia
in ground-based observations. 
The follow-up spectroscopy can be attempted on the Keck Low-Resolution
Imaging Spectrometer (LRIS). Assuming that we use the 300 grooves/mm
grating which provide dispersions of 4.99$\,\AA$ per 48$\,\mu$m 
(2 pixels on the CCD), we find the exposure time for 0.5 arcsec seeing
to be
\ba
t&=&28 \,{\rm hours}\, \left(\frac{S/N}{3}\right)^2 \left(\frac{W}{1\, 
{\rm arcsec} }
\right)\, 10^{0.4 [ 2(I_{AB}-26)-(I^{sky}_{AB}-21)]},\nonumber\\
&=&4.44 \,{\rm hours}\, \left(\frac{S/N}{3}\right)^2 \left(\frac{W}{1\, 
{\rm arcsec} }
\right)\, 10^{0.4 [ 2(I_{AB}-25)-(I^{sky}_{AB}-21)]},\nonumber\\
&=&12.33 \,{\rm hours}\, \left(\frac{S/N}{5}\right)^2 \left(\frac{W}{1\, 
{\rm arcsec} }
\right)\, 10^{0.4 [ 2(I_{AB}-25)-(I^{sky}_{AB}-21)]},
\ea
where $W$ is the width of the slit, and $I^{sky}_{AB}$ is the 
sky brightness in the I band in units of mag/arcsec$^2$.
Note that the Si \Roman{ctr} absorption feature at $\sim 4000\,\AA$
is not deep enough to be useful in a very noisy ($S/N=3$) spectrum.
Clearly, spectroscopic follow-up of the $z\sim 1.5$ ($I_{AB}\sim 26$)
SNe Ia will require substantial observational resources; NICMOS or NGST 
will be more suitable for the spectroscopy of the SNe Ia at the highest
redshifts discovered from the ground.
The Keck LRIS can be used to obtain the spectra of the
SNe Ia at more modest redshifts ($z\sim 1.3$, $I_{AB}\sim 25$).

A dedicated 4 meter telescope with a square degree field of view
can be used to conduct other important scientific projects compatible 
with the SN pencil beam survey, such as QSOs, Kuiper belt objects, 
and in particular, weak lensing and
the search of gamma-ray burst (GRB) afterglows. 

Weak lensing is a powerful tool in mapping the mass distribution
in the universe. The large field of view and the depth of a pencil
beam survey would be ideal for weak lensing measurements of 
field galaxies, which can be used to constrain the large scale structure
in the universe.

GRBs are perhaps the most energetic astrophysical events in the universe.
Currently, there are many competing theories to explain GRBs.
GRB afterglows contain valuable information, a statistically significant
sample of GRB afterglows can provide strong constraints on the GRB theory. 
If beaming is involved in GRBs, we expect most of the GRB afterglows 
not to be associated with observable bursts. Schmidt et al. (1998) 
have observed a few optical transients which were too short in duration 
to be SNe; these could be GRB afterglows. Since present observational
data seem to indicate that GRBs have associated host galaxies,
the detection of host galaxies associated with short optical
transients would support the interpretation of the latter as
GRB afterglows. Since the GRB host galaxies are typically fainter
than $R=25$, a pencil beam survey would be ideal in detecting these
host galaxies of candidate GRB afterglows.

%\newpage

\section{Summary and discussion}

We have proposed a pencil beam survey of SNe Ia which can yield
from tens to hundreds of SNe Ia per 0.1 redshift interval for $z$
up to 1.5, which would enable the quantitative consideration of
the systematic uncertainties of SNe Ia as standard candles,
in particular, luminosity evolution and gravitational lensing.
Using the Perlmutter et al. ``batch''
search technique repetitively over the same field in the sky,
the pencil beam survey would be efficient in the discovery of
SNe at $z\ga 1$ by allowing the comparison of the latest frame of images 
(which may consist of combined data from successive nights)
with all previous frames.

The direct product of such a survey is the SN number count
as a function of redshift (see \S 3), which is a combined measure of
the cosmological parameters and the SN rate. When the
measurements of the cosmological parameters are applied as priors
to the number count, we obtain a direct measure of the
SN rate, which is a key parameter in the formation and evolution
of galaxies. The non-type-Ia SNe discovered by
the pencil beam survey may be comparable to the type Ia SNe in number.

The most important and straightforward application of the data from the SN 
pencil beam survey is to reduce the gravitational lensing
noise in a SN Ia Hubble diagram via flux averaging. 
We have simulated SN Ia luminosities by adding weak lensing noise
(using empirical fitting formulae given by Wang 1999a) and 
scatter in SN Ia absolute magnitudes
to standard candles placed at random redshifts.
We have shown that flux-averaging is powerful in reducing the
combined noise of gravitational lensing and SN Ia absolute magnitude
scatter (see \S 2).
Because of the non-Gaussian nature of the luminosity distribution 
of SNe Ia at given $z$ due to weak lensing, the large number of SNe Ia 
in a given redshift interval at high $z$ is essential for the proper 
modeling and removal of the gravitational lensing effect. 
The SN Ia luminosity distribution in each redshift interval 
can be used to constrain the cosmological model 
(in particular, the fraction of matter in compact objects) by
comparison with predictions of numerical ray-shooting. (\cite{Holz98})   

The completeness of the SN Ia sample determines
the effectiveness of the removal of gravitational lensing noise from
the SN Ia Hubble diagram and the amount of information contained
in the SN Ia luminosity distribution in each redshift interval.
Note that the magnitude limit of the survey can lead to observational
bias against the most distant demagnified SNe, therefore, the SNe
which are close to the magnitude limit of the survey should not be
used to probe cosmology in the manner described in this paper. 
To ensure the maximum usefulness of the data, scrupulous
attention will have to be paid to photometric calibration, 
uniform treatment of nearby and distant samples, and an effective 
way to deal with reddening. (\cite{Riess98})

Note that although we can remove or reduce the effect of gravitational 
lensing on the SN Ia Hubble diagram,
other systematics can affect the observed luminosity of SNe Ia.
For example, grey dust,
an evolution of the reddening law, or evolution in
the peak absolute luminosity of SNe Ia. While it is
possible to constrain grey dust and determine dust evolution through 
multi-band photometry at significantly different redshifts,
the dimming with $z$ in peak absolute luminosity of SNe Ia is degenerate
with the effect of low matter density. 
(\cite{Aguirre99,Riess99,DLW99,Wang99b})
Evolution will remain a caveat
in the usage of SNe Ia as cosmological standard candles, unless one
can somehow correct for the effect of evolution.
It is critical to obtain up to hundreds of SNe Ia at $z>1$ to
constrain luminosity evolution, because we expect the luminosity
evolution and low matter density to affect the
distance modulus of SNe Ia through
different functionals of $z$, which should become distinguishable at $z>1$.

We have proposed a survey cutoff of $z=1.5$ mainly for two reasons.
First, going to higher redshift makes obtaining the spectra
of the SNe (which are needed to distinguish different types of SNe)
practically impossible from the ground; even obtaining the spectra of
$z\sim 1.5$ SNe Ia may prove impractical from the ground (as it would require
several clear nights per spectrum on the Keck, see \S 4).
Observers have already demonstrated that SNe Ia up to $z=1$ can be
found in the ongoing searches. (\cite{Goo95,Garna98,Perl99})
At $z\sim 1.5$, the SNe Ia should be $\sim 2.5$ magnitudes
fainter than at $z\sim 1$, it should be possible to discover them
on the ground through deep imaging, and the follow-up spectroscopy 
can be done on NICMOS or NGST (see \S 4). 
Second, predicted rest-frame SN Ia rate per comoving volume
as function of redshift seems to peak at $z\sim 1$. (\cite{Yung98,Sadat98})
A pencil beam survey of SNe up to $z=1.5$ will enable the accurate
determination of the SN rate as function of redshift in the redshift
region important for studying cosmic star formation rate and the
SN Ia progenitor models. Although the Next Generation Space Telescope
can detect SNe at as high redshifts as they exist (\cite{Stockman98}), 
the estimated rate of detection is of order 20 SNe II per 4$\times$4 
arcmin$^2$ field per year in the interval $1<z<4$ (\cite{Madau98b}),
and the detection rate of SNe Ia is likely smaller.
Thus a ground-based pencil beam survey of SNe is essential
to complement the space-based SN searches.

The most challenging aspect of the SN pencil beam survey is obtaining
spectra for the SNe Ia at redshifts close to 1.5. Instead of waiting 
for future space equipments, we may find innovative ways of 
obtaining spectra from the ground.
The referee has pointed out that since we can find multiple SNe 
at the same time on the same one square degree field (the number of SNe 
depends on cosmology), it may be possible to get multiple spectra 
at once with fibers.

The nominal numbers we have used for the proposed SN pencil beam survey,
a one square degree field, a depth of $z=1.5$, and an effective 
observation duration of one year (which is equivalent to several years of
actual observation), are optimistic but not implausible.
The large sky coverage and the long effective observation
duration will probably require a large consortium of existing and new
SN search teams through, for example, a dedicated
4 meter telescope which can be used at the same time for other
important observational projects compatible with the SN pencil beam survey
(see \S 4).
The goal of going up to $z=1.5$ in spectroscopy will
require support from the Keck LRIS and NICMOS/NGST.
We conclude by noting that a SN pencil beam survey 
can yield enormous scientific return.
The observational efforts directed towards a SN pencil beam
survey should be very rewarding.

\acknowledgements{\centerline{\bf Acknowledgements}}

It is a pleasure for me to thank Joachim Wambsganss for generously providing 
unpublished magnification distributions; 
Zeljko Ivezic and Todd Tripp for explaining technical
details concerning photometry and spectroscopy;
Saul Perlmutter for communicating details
of the current supernova search by the Supernova Cosmology Project;
Ed Turner for a careful reading of a draft of the manuscript
and for helpful comments, the referee for encouraging and useful
comments; Christophe Alard, Jim Gunn, David Hogg, 
Rocky Kolb, Robert Lupton, Michael Strauss, and Tony Tyson for 
helpful discussions.

%\end{document}

\clearpage
\setcounter{figure}{0}
\figcaption[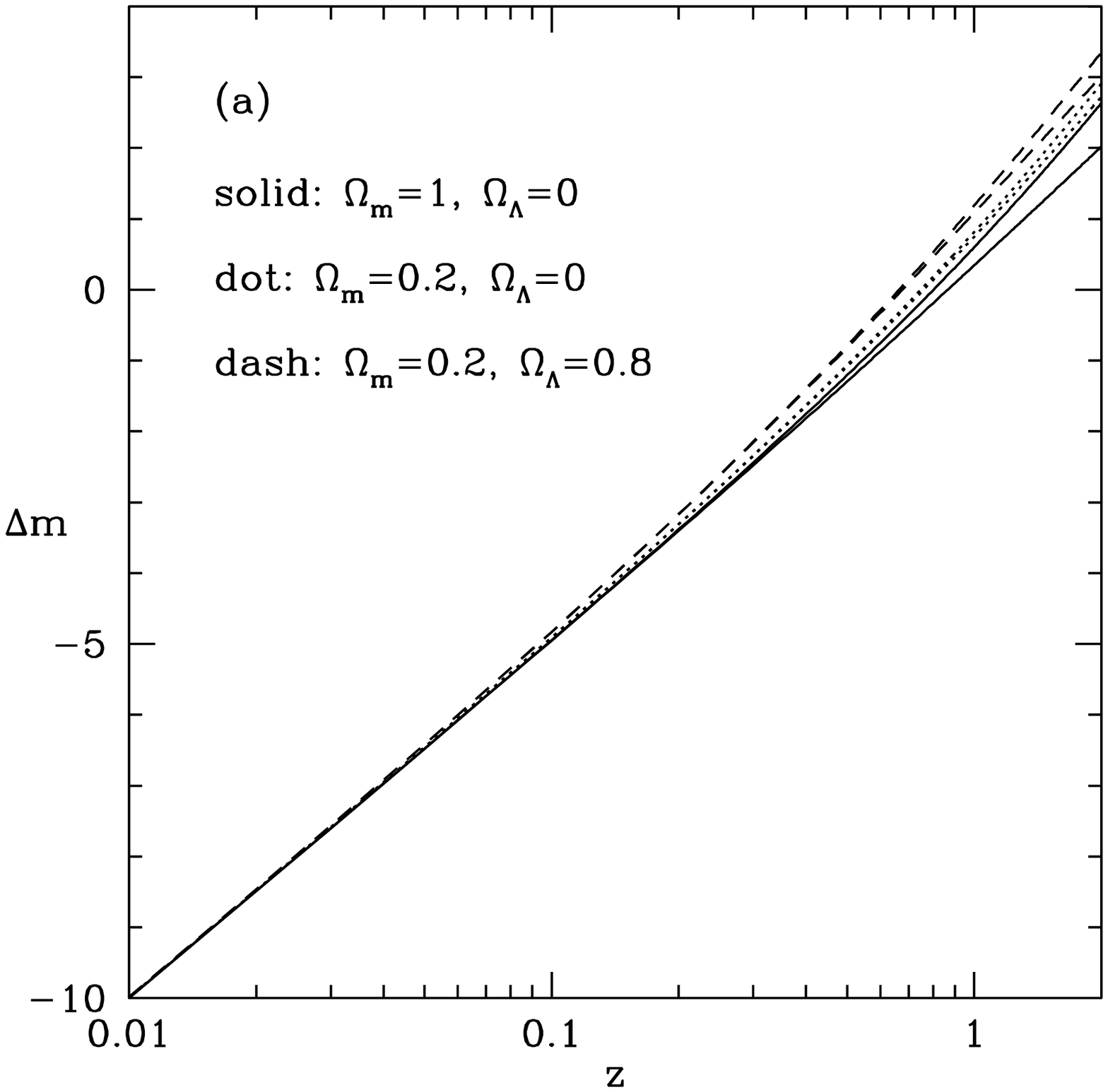]{
(a) Magnitude versus redshift for three cosmological models,
SCDM ($\Omega_m=1$, $\Omega_{\Lambda}=0$), 
OCDM ($\Omega_m=0.2$, $\Omega_{\Lambda}=0$),
and $\Lambda$CDM ($\Omega_m=0.2$, $\Omega_{\Lambda}=0.8$).
For each cosmological model, the upper curve represents the completely
clumpy universe (empty beam, $\tilde{\alpha}=0$), 
while the lower curve represents the completely smooth universe 
(filled beam, $\tilde{\alpha}=1$).
(b) The same models relative to smooth OCDM (filled beam, 
$\tilde{\alpha}=1$), the middle curve for each model represents 
a universe with half of the matter smoothly distributed 
(half-filled beam, $\tilde{\alpha}=0.5$). }

\figcaption[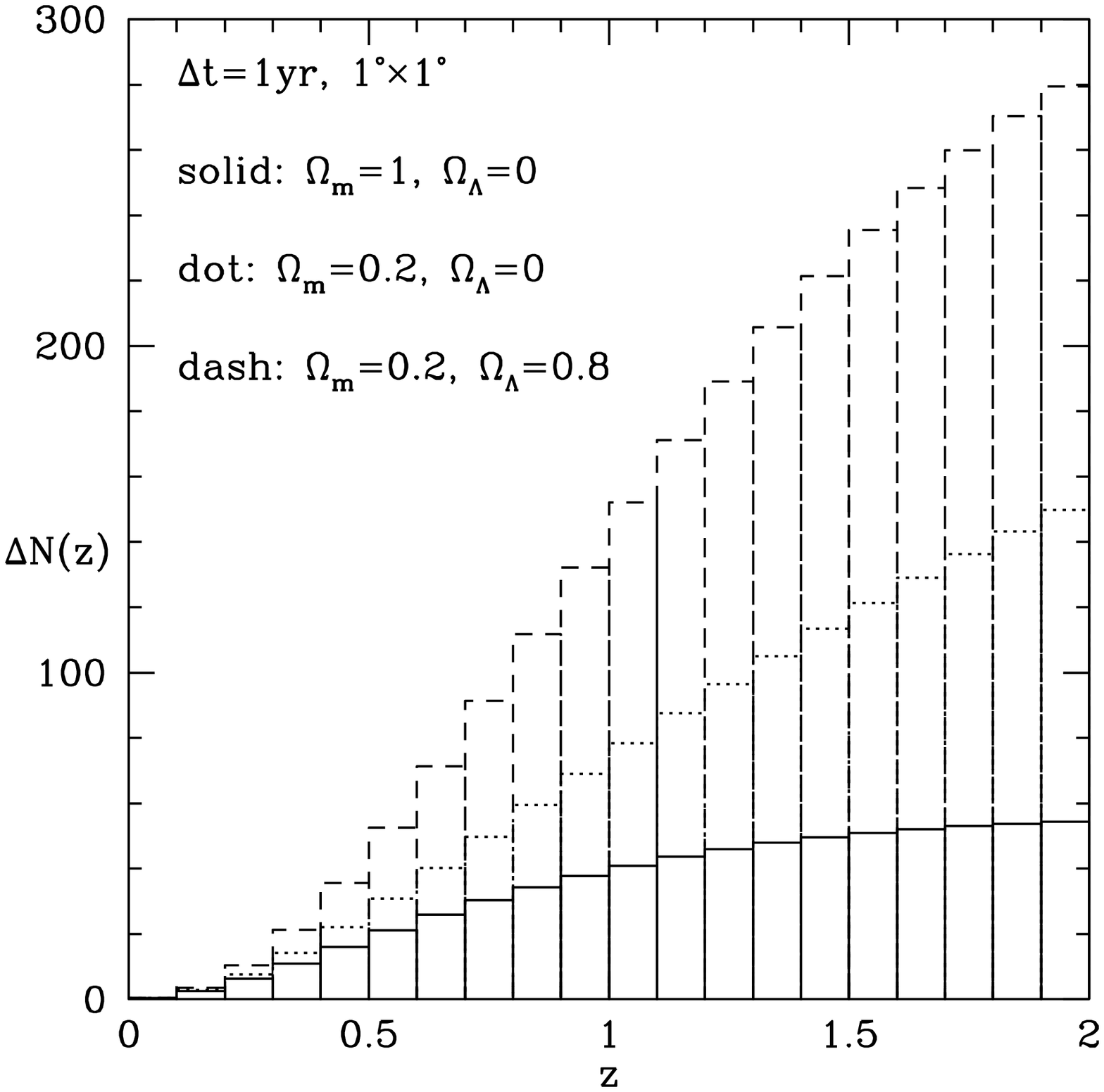]{
The number of SNe Ia expected per 0.1 redshift interval 
as function of redshift for three cosmological models,
SCDM ($\Omega_m=1$, $\Omega_{\Lambda}=0$), 
OCDM ($\Omega_m=0.2$, $\Omega_{\Lambda}=0$),
and $\Lambda$CDM ($\Omega_m=0.2$, $\Omega_{\Lambda}=0.8$),
for a one square degree field, and an effective observation 
duration of one year.}

\figcaption[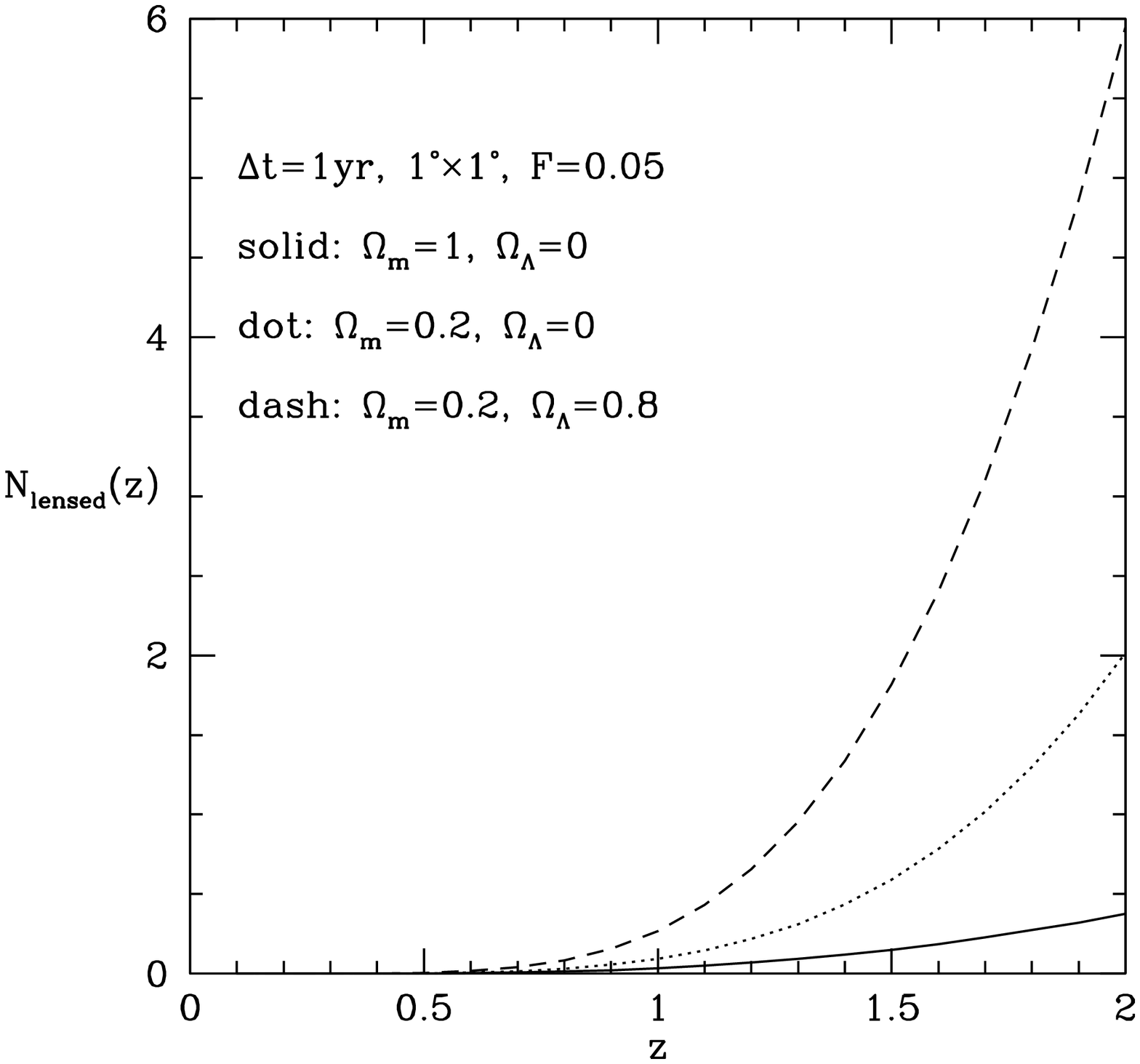]{
The number of strongly lensed SNe as function of
survey depth $z$, for the same cosmological models as in Fig.1 and Fig.2,
for a one square degree field, and an effective observation duration 
of one year.}

\figcaption[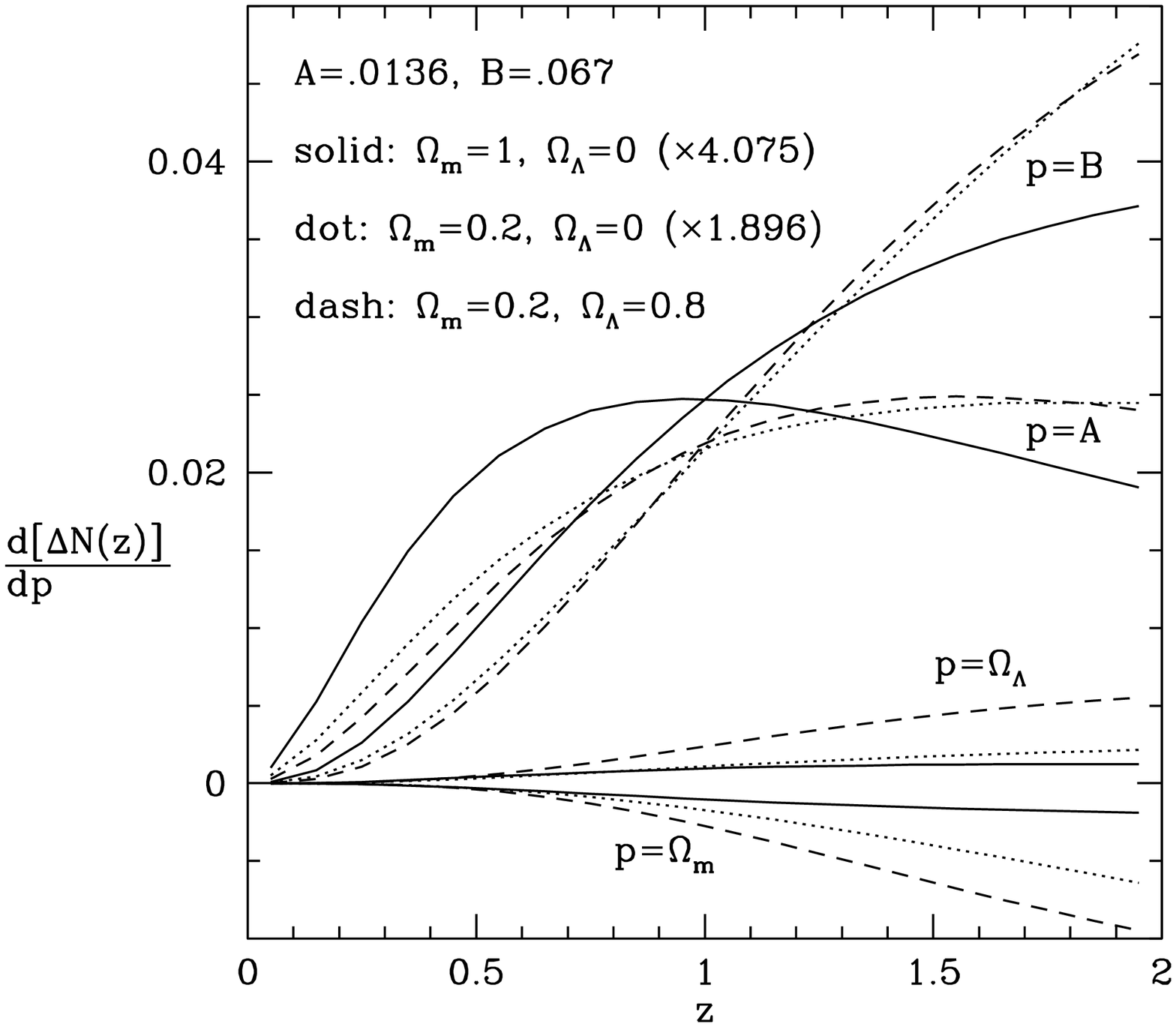]{
The parameter dependence of the SN Ia number count per 0.1 
redshift interval as function of redshift. Note that the dependence
of the SN number count on the SN rate is parametrized by $A$ and $B$.}

\clearpage

\setcounter{figure}{0}
\plotone{fig1a.eps}
\figcaption{
(a) Magnitude versus redshift for three cosmological models,
SCDM ($\Omega_m=1$, $\Omega_{\Lambda}=0$), 
OCDM ($\Omega_m=0.2$, $\Omega_{\Lambda}=0$),
and $\Lambda$CDM ($\Omega_m=0.2$, $\Omega_{\Lambda}=0.8$).
For each cosmological model, the upper curve represents the completely
clumpy universe (empty beam, $\tilde{\alpha}=0$), 
while the lower curve represents the completely smooth universe 
(filled beam, $\tilde{\alpha}=1$).}

\setcounter{figure}{0}
\plotone{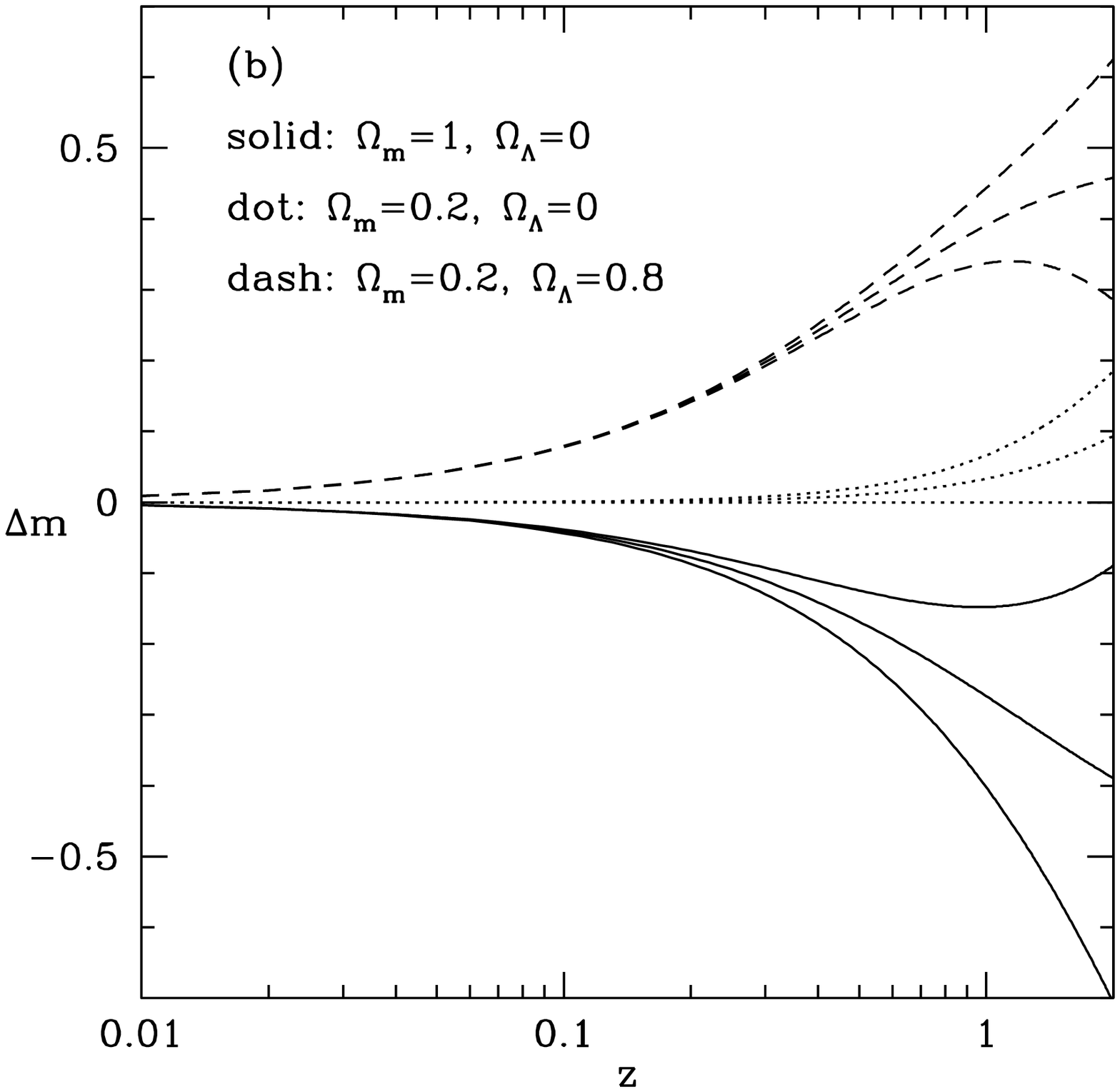}
\figcaption[fig1b.eps]{
(b) The same models relative to smooth OCDM (filled beam, 
$\tilde{\alpha}=1$), the middle curve for each model represents 
a universe with half of the matter smoothly distributed 
(half-filled beam, $\tilde{\alpha}=0.5$). }

\plotone{fig2.eps}
\figcaption[fig2.eps]{
The number of SNe Ia expected per 0.1 redshift interval 
as function of redshift for three cosmological models,
SCDM ($\Omega_m=1$, $\Omega_{\Lambda}=0$), 
OCDM ($\Omega_m=0.2$, $\Omega_{\Lambda}=0$),
and $\Lambda$CDM ($\Omega_m=0.2$, $\Omega_{\Lambda}=0.8$),
for a one square degree field, and an effective observation 
duration of one year.}

\plotone{fig3.eps}
\figcaption[fig3.eps]{
The number of strongly lensed SNe as function of
survey depth $z$, for the same cosmological models as in Fig.1 and Fig.2,
for a one square degree field, and an effective observation duration 
of one year.}

\plotone{fig4.eps}
\figcaption[fig4.eps]{
The parameter dependence of the SN Ia number count per 0.1 
redshift interval as function of redshift. Note that the dependence
of the SN number count on the SN rate is parametrized by $A$ and $B$.}

\end{document}